\documentclass[%
 nofootinbib,reprint,
 amsmath,amssymb,
 aps
]{revtex4-2}

\usepackage{graphicx}% Include figure files
\usepackage{subfig}
\usepackage{tabularx}% Include figure files
\usepackage{dcolumn}% Align table columns on decimal point
\usepackage{bm}% bold math
\usepackage{mathrsfs}
\usepackage{float}
\usepackage{url}
\newcommand{\nm}{\textrm{nm}}

\newcommand{\um}{\textrm{\textmu m}}
\newcommand{\dB}{\textrm{dB}}

\newcommand{\twoline}[2]{\begin{tabular}{@{}c@{}}#1\\#2\end{tabular}}

\begin{document}

\title{Waveguide-Integrated Mid-Infrared Photodetection using Graphene on a Scalable Chalcogenide Glass Platform}

\author{Jordan Goldstein}
\author{Hongtao Lin}
 \altaffiliation[Present address: ]{College of Information Science and Electronic Engineering, Zhejiang University, Hangzhou, China.}
\author{Skylar Deckoff-Jones}
\author{Marek Hempel}
\author{Ang-Yu Lu}
% maybe \author{Ji-Hoon Park}
\affiliation{%
Massachusetts Institute of Technology, 77 Massachusetts Ave, Cambridge MA 02139, USA
}%
\author{Kathleen A. Richardson}
\affiliation{University of Central Florida, 4000 Central Florida Blvd, Orlando, FL 32816}
\author{Tom\'as Palacios}
\author{Jing Kong}
\author{Juejun Hu}
\author{Dirk Englund}%
 \email{Corresponding author: englund@mit.edu}
\affiliation{%
Massachusetts Institute of Technology, 77 Massachusetts Ave, Cambridge MA 02139, USA
}%

\date{\today}

\begin{abstract}
% Make starting sentence broader
% "An essential challenge in..."
% START with what I want to acheive
% CONTINUE with the solution
% "Here we intoduce a new platform" I.E. don't just make it sound like we shifted the wavelength up.
% "Through these innovations, we acheive..."
% Set grand challenges
% Give state of the art
% Show you are making a profound change
% What has it done for us? (I.E. quantitative result)
% "Recently black phosphorus" here we show that our system...

% CITATIONS HAVE BEEN REMOVED FROM ABSTRACT TO CONFORM TO NATCOMM PUBLICATION GUIDELINES
The development of compact and fieldable mid-infrared (mid-IR) spectroscopy devices represents a critical challenge for distributed sensing with applications from gas leak detection to environmental monitoring. Recent work has focused on mid-IR photonic integrated circuit (PIC) sensing platforms and waveguide-integrated mid-IR light sources and detectors based on semiconductors such as PbTe, black phosphorus and tellurene. However, material bandgaps and reliance on SiO$_2$ substrates limit operation to wavelengths $\lambda\lesssim4\,\um$. Here we overcome these challenges with a chalcogenide glass-on-CaF$_2$ PIC architecture incorporating split-gate photothermoelectric graphene photodetectors. Our design extends operation to $\lambda=5.2\,\um$ with a Johnson noise-limited noise-equivalent power of $1.1\,\mathrm{nW}/\mathrm{Hz}^{1/2}$, no fall-off in photoresponse up to $f = 1\,\mathrm{MHz}$, and a predicted 3-dB bandwidth of $f_{3\dB}>1\,\mathrm{GHz}$. This mid-IR PIC platform readily extends to longer wavelengths and opens the door to applications from distributed gas sensing and portable dual comb spectroscopy to weather-resilient free space optical communications.
\end{abstract}

\maketitle

\section*{}

\begin{figure}[tb]
    \centering
    \includegraphics[width=\columnwidth]{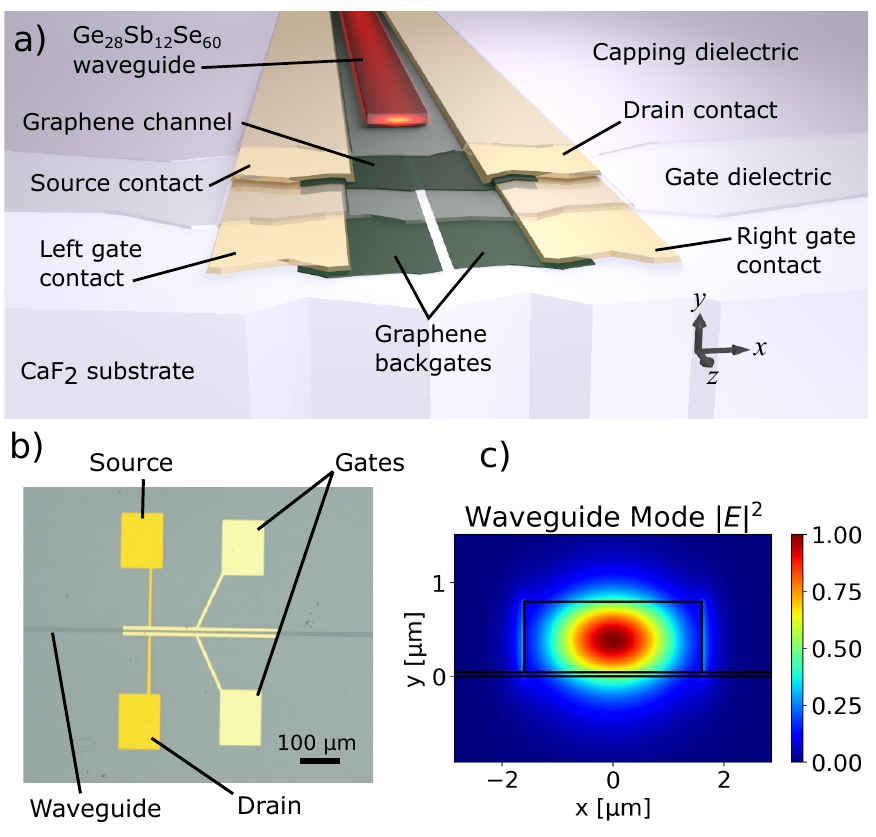}
    \caption{a) Illustration of the device cross-section perpendicular to the waveguide axis. The optical mode supported by the GSSe waveguide evanscently couples to and is absorbed by the graphene channel, which is gated by two graphene back-gates to induce a \emph{p-n} junction. b) Optical image of the device depicting source, drain and gate contact pads. c) Depiction of the optical guided mode at $\lambda=5.2\,\um$.}
    \label{devicedesign}
\end{figure}

\begin{figure*}[tb]
    \centering
    \includegraphics[width=\textwidth]{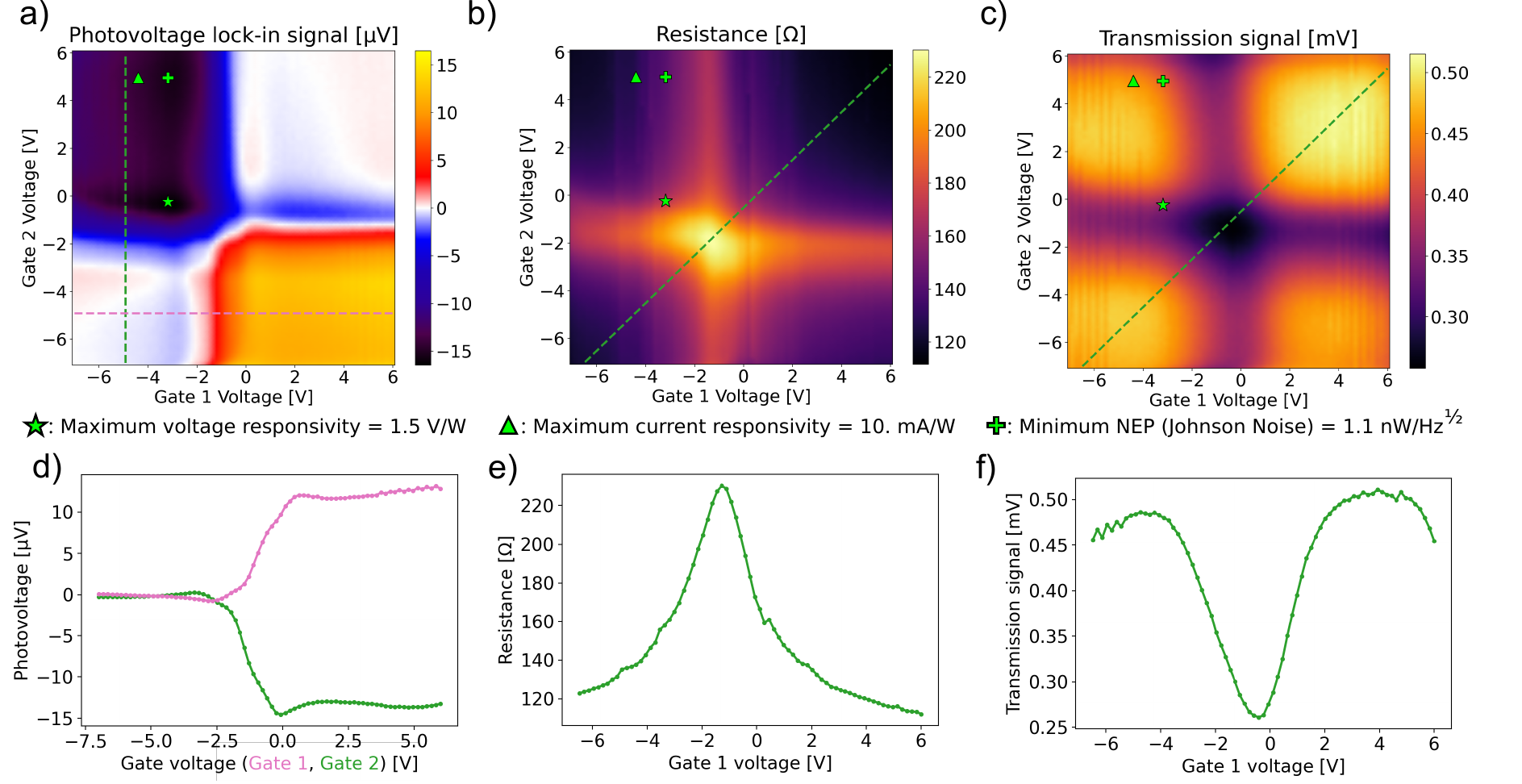}
    \caption{a) Measured zero-bias photovoltage produced by the device as a function of the two gate voltages. b) Total device resistance as a function of the two gate voltages. c) Lock-in signal reflecting power measured by an InAsSb photodetector at the focal point of our output facet collection lens, used to monitor transmission of the device as a function of the gate voltages. The power-normalized transmitted is plotted in Supplementary Fig. 3b. d, e, f) Plots of line sections indicated with dashed lines in figures a, b, and c, respectively.}
    \label{rawmeas}
\end{figure*}

Mid-IR absorption spectroscopy is a critical tool for chemical sensing and analysis, especially for inert gases that evade detection by chemical reaction-based sensors. Many such gases derive their inertness from halogenated chemistries and thus exhibit global warming potential due to carbon-halogen stretching modes resonant in the thermal IR\cite{irspectroscopy,cfcs}. To facilitate sensor deployment for greenhouse gas leak detection and other chemical sensor application areas, there exists a strong need to transition from co-packaged discrete components to compact and chip-integrated sensors.%, augmenting the importance of rapid detection of leaks of these gases. Current infrared gas sensors use co-packaged discrete optics, the size and cost of which precludes broad, networked deployment around chemical plants, electrical utility sites, and other locations liable to greenhouse gas leaks, hampering sensitive and responsive leak detection.

To address this challenge, mid-IR photonic integrated circuit platforms have been investigated to reduce optical gas sensors to the size of a chip. Recent work has demonstrated integrated optical methane\cite{methaneslot} and volatile organic compound\cite{ptlin} sensing, but required coupling to off-chip sources and detectors. However, integrating the detector on-chip is more compact and can improve sensitivity by reducing the volume of active material able to generate thermal noise. %, as the noise-equivalent power (NEP) of an infrared photodetector is typically proportional to the square root of its area\cite{rogalski}. %Shrinking the detector to the size of an optical waveguide mode maximizes NEP, allowing a better power consumption versus signal-to-noise ratio trade-off. 
Su et al. achieved integration of a PbTe photoconductor and demonstrated methane sensing at a wavelength of $\lambda=3.31\,\um$\cite{petersu}, but their platform is limited to $\lambda\lesssim4\,\um$ due to absorption in the SiO$_2$ substrate\cite{anu} and by PbTe's absorption cutoff\cite{rogalski}. Waveguide-integrated detectors based on narrow-gap 2D materials black phosphorus\cite{bp} and tellurene\cite{skylartellurene} have also been demonstrated, but they too are bandgap-limited to $\lambda\lesssim4\,\um$.

Here we exceed the wavelength limit of previous demonstrations using graphene-based detectors on an extended-transparency waveguide platform. While graphene integrated detectors have shown promise at telecom wavelengths\cite{plasmonic}, the material's advantages are magnified further at longer wavelengths due to the thermal nature of the photothermoelectric (PTE) response mechanism\cite{gabor,jsong} and due to the impact of optical plasmon scattering at short wavelengths\cite{carriercarrier}. Integrated photodetection with graphene has been demonstrated at wavelengths up to $3.8\,\um$\cite{anu} and with chalcogenide glass waveguides\cite{hongtao}, but on SiO$_2$ platforms. To access longer wavelength operation and achieve good sensitivity at zero bias, we introduce a Ge$_{28}$Sb$_{12}$Se$_{60}$ (GSSe)-on-CaF$_2$ waveguide platform supporting gated PTE-based graphene photodetectors.

Figs. \ref{devicedesign}a and \ref{devicedesign}b illustrate the platform and photodetector design. The device consists of a single-mode GSSe waveguide fabricated on top of a $5.4\,\um$ wide by $300\,\um$ long, CVD-grown graphene channel, flanked on either side by source and drain contacts. Beneath the graphene channel are pair of CVD graphene back-gates, separated by a $400\,\mathrm{nm}$ gap and used to electrostatically induce a \emph{p-n} junction along the center of the channel. We use HfO$_2$ as the gate dielectric and as an airtight capping layer. The device is fabricated on a CaF$_2$ substrate, transparent up to $\lambda=8\,\um$. Fig. \ref{devicedesign}c depicts the resulting waveguide mode at $\lambda=5.2\,\um$. %Light enters and exits the chip via a pair of cleaved waveguide facets on opposing sides of the chip, which are laterally offset by 5 mm to reduce the amount of stray input light passing through the aperture stop of our collection optics.

We use lock-in measurement to characterize our detectors, focusing light from a $\lambda=5.2\,\um$ QCL source into our chip's input facet. Light exiting the chip is focused onto an InAsSb photodetector and amplified for transmission measurement. Supplementary Fig. 1a depicts this beam-path in more detail. We operate the device under zero bias voltage to avoid introducing electronic shot noise and to prevent channel conductivity fluctuations from manifesting as $1/f$ noise\cite{flickernoise}. For the following low-frequency measurements we use a lock-in amplifier to measure the photovoltage directly with no preamplification.

Figs. \ref{rawmeas}a, b and c plot the photovoltage, resistance, and transmission lock-in signals versus both gate voltages for one such photodetector (``Device A''). Here, we modulate the $\lambda=5.2\,\um$ QCL source at $3.78\,\mathrm{kHz}$ with a guided ``on'' power of $11\,\textrm{\textmu W}$ at the detector input. From our photovoltage and resistance maps, alongside the power and waveguide loss calibrations described in Supplementary Section A, we infer the gate voltage pairs that optimize the voltage responsivity, current responsivity, and NEP with respect to Johnson noise, indicated with green markers in Fig. \ref{rawmeas}. For these, we arrive at $1.5\,\mathrm{V}/\mathrm{W}$, $10.\,\mathrm{mA}/\mathrm{W}$, and $1.1\,\mathrm{nW}/\mathrm{Hz}^{1/2}$, respectively. The observed photovoltage gate map indicates a PTE response mechanism, evidenced by the six-fold sign change pattern around the graphene channel's charge neutral point\cite{gabor}. Figs. \ref{rawmeas}d, e and f show line slices of the voltage maps as indicated by the dashed lines in Figs. \ref{rawmeas}a, b and c of the same color. Fig. \ref{rawmeas}d, in particular, highlights the changes in slope associated with PTE-based detectors\cite{gabor}.

\begin{figure}[tb]
    \centering
    \includegraphics[width=0.80\columnwidth]{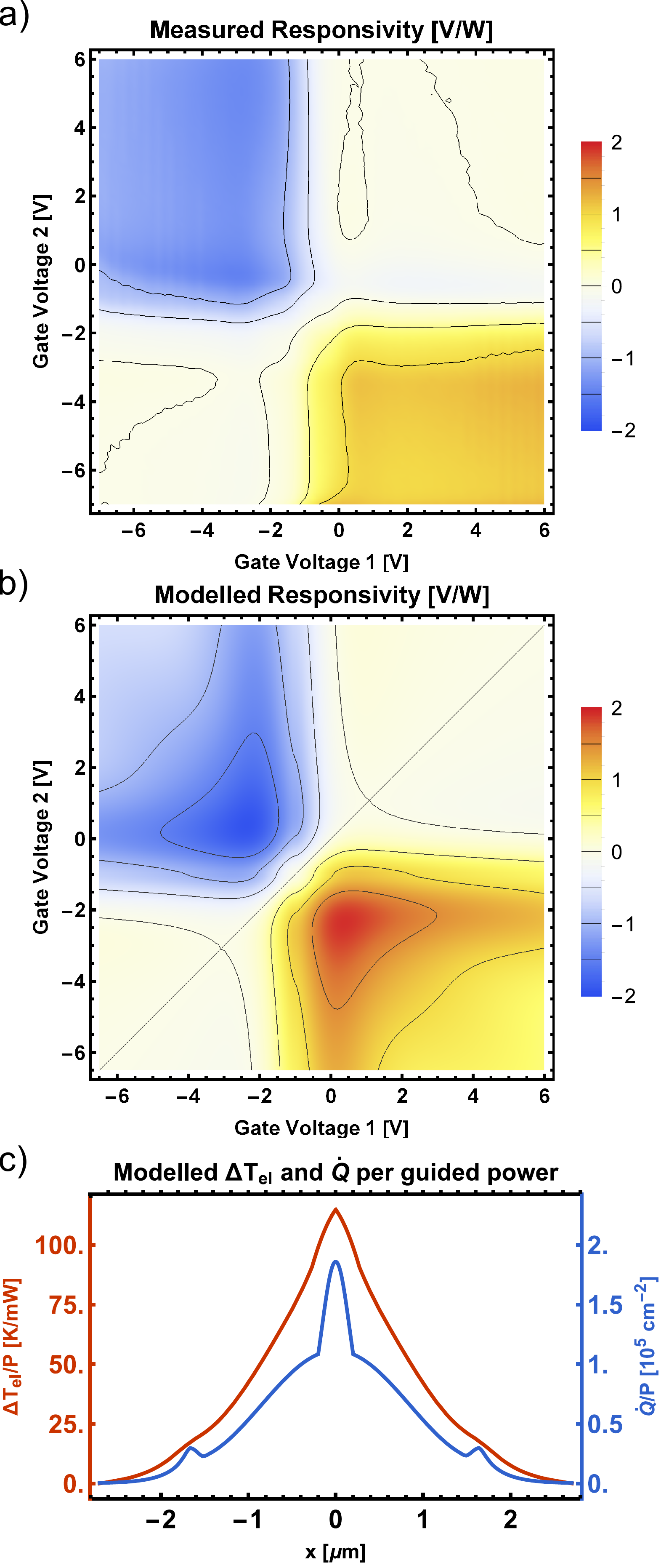}
    \caption{a, b) Contour plots of the a) measured and b) modelled responsivity maps of our device, evaluated with $\tau_\mathrm{DC}=3.5\,\mathrm{fs}$, $\tau_\mathrm{IR}=40\,\mathrm{fs}$, $\sigma_\mathrm{n}=2\times10^{12}\,\mathrm{cm}^{-2}$, $\tau_\mathrm{eph}=50\,\mathrm{ps}$, and $\alpha_\mathrm{e}=2.5\,\mathrm{mm}^{-1}$. %Except for $\tau_\mathrm{eph}$ and $\eta$, the graphene quality and waveguide loss parameters used to model the responsivity are derived from fitting the device resistance and transmittance maps. 
    c) Electron temperature increase $\Delta T_\mathrm{el}$ and absorbed optical power per area $\dot{Q}$ profiles in the graphene channel per guided optical power at gate voltages of $\{-2.35\,\mathrm{V},\,0.35\,\mathrm{V}\}$, chosen to maximize the modelled photoresponse, and other parameters as above.}
    \label{comparison}
\end{figure}

To confirm our understanding of device operation and elucidate the prospects for performance improvement, we apply the formalism introduced in Song et al.\cite{jsong} to calculate the electronic temperature distribution and Seebeck photovoltage in the graphene channel under illumination. Figs. \ref{comparison}a and \ref{comparison}b compare our measured and modelled voltage responsivities using calculations described in the Methods section. The performance of our device depends on several fitting parameters, whose definitions and approximate values (derived from our measured data) we provide in Table \ref{params}. We describe our fitting process in Supplementary Section C. Critically, all features of the modelled responsivity map in Fig. \ref{comparison}b up to an overall scale factor from $\tau_\mathrm{eph}$ are established \emph{a priori} from fitting parameters extracted from the device transmittance and resistance maps, with only $\tau_\mathrm{eph}$ obtained by matching the scales of the measured and modelled responsivities. The resemblance between Figs. \ref{comparison}a and \ref{comparison}b thus reflects the validity of our PTE model and is not due to over-fitting. In Fig. \ref{comparison}c, we plot the solution to Eqn. \ref{ode}, $\Delta T_\mathrm{el}(x)$, as well as the source term $\dot{Q}(x)$. The thermal transport model predicts that $9\,\textrm{\textmu W}$ of guided power raises the temperature of the graphene channel's electron gas by as much as $1\,\mathrm{K}$ along the center of the device.

\begin{table}[tb]
\centering
\caption{Device parameters and approximate values}
%\subfloat[Parameter definitions]{
\begin{tabular}{|c|c|c|}
\hline
 \rule{0pt}{10pt}$\tau_\mathrm{DC}$ & Drude scattering time at DC &$\approx3.5\,\mathrm{fs}$\\\hline
 \rule{0pt}{10pt}$\tau_\mathrm{IR}$ & Drude scattering time at IR &$30\text{--}50\,\mathrm{fs}$\\\hline  
 \rule{0pt}{16pt}$\sigma_\mathrm{n}$ & 
 \begin{tabular}{@{}c@{}}Standard deviation of native carrier\\ concentration due to spatial inhomogeneity \end{tabular} &
 \begin{tabular}{@{}c@{}}$1.5\text{--}2.5\,\times$\\$10^{12}\,\mathrm{cm}^{-2}$\end{tabular} \\\hline
 \rule{0pt}{10pt}$E_\mathrm{Fc}$ & Native Fermi level of graphene channel &$\approx0.17\,\mathrm{eV}$\\\hline
 \rule{0pt}{10pt}$E_\mathrm{Fg}$ & Native Fermi level of graphene gates &$\approx0.48\,\mathrm{eV}$\\\hline
 \rule{0pt}{10pt}$\tau_\mathrm{eph}$ & Electron-phonon cooling time &$\approx50\,\mathrm{ps}$\\\hline
 \rule{0pt}{10pt}$\alpha_\mathrm{e}$ & Excess light attenuation within device&$2\text{--}3\,\mathrm{mm}^{-1}$\\\hline
\end{tabular}
\label{params}
\end{table}

\begin{figure}[tb]
    \centering
    \includegraphics[width=\columnwidth]{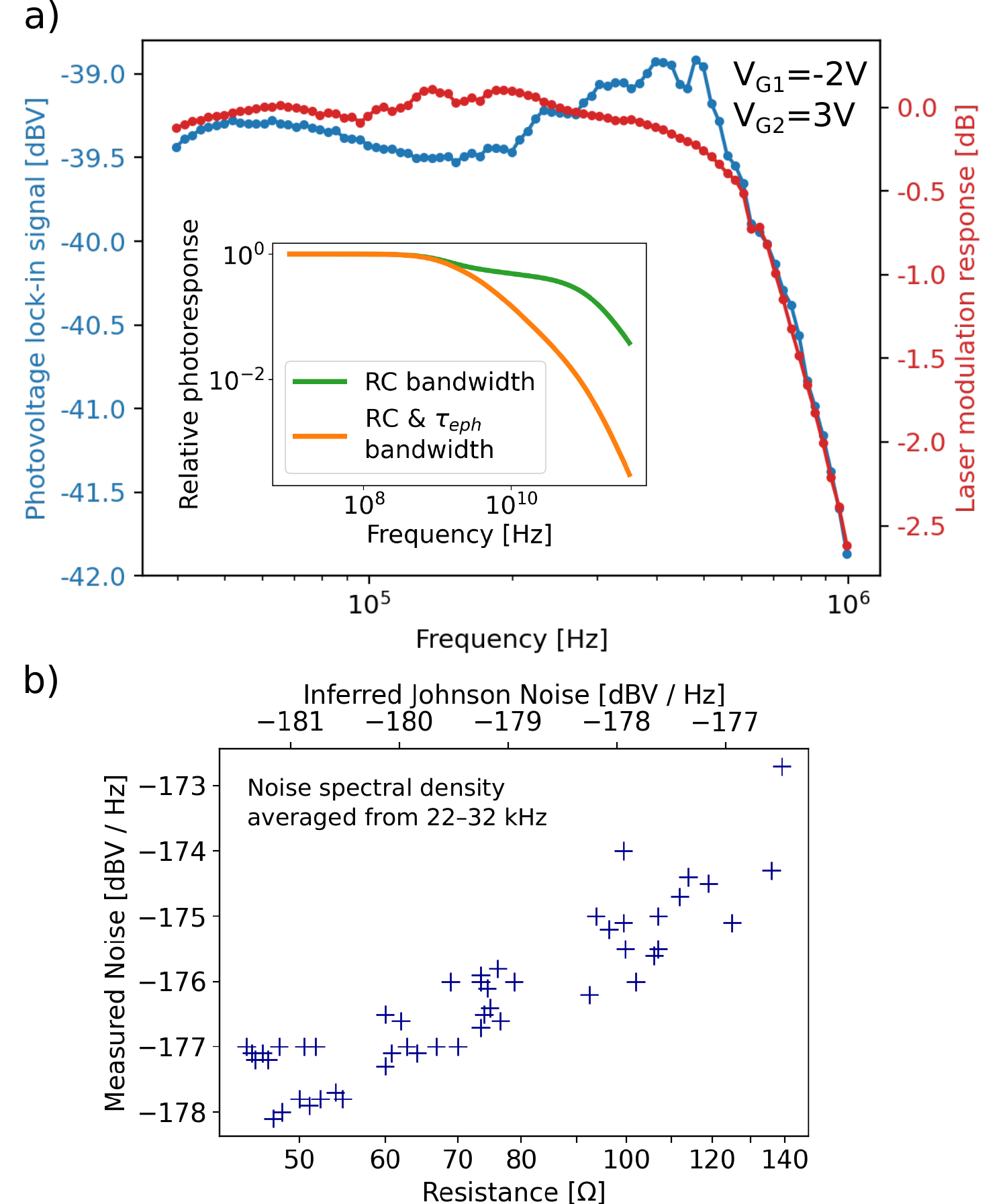}
    \caption{a) Comparison of the frequency response of our photodetector with that of the laser current modulation itself. The consistency between the two indicates that the photodetector frequency response exceeds 1 MHz. Inset: Simulated GHz-range photodetector frequency response, with and without considering the impact of the electron-phonon cooling time $\tau_\mathrm{eph}$. b) Measured noise spectral density versus resistance and corresponding Johnson noise spectral density of Device B, without illumination, for the 49 pairs of gate voltages $\{V_\mathrm{g1},V_\mathrm{g2}\}$ where each $V_\mathrm{gn}$ is varied from $-6\,\mathrm{V}$ to $6\,\mathrm{V}$ in steps of $2\,\mathrm{V}$. Measurement was performed at $T=293\,\mathrm{K}$.}
    \label{freqresp}
\end{figure}

Current modulation of our QCL source permits frequency response measurements up to its modulation bandwidth of $1\,\mathrm{MHz}$. To account for the modulation response of our laser, we measure the photovoltage of Device A alongside that of a fast InAsSb photodiode. The comparison shown in Fig. \ref{freqresp}a indicates that our device is faster than our laser's modulation bandwidth. We thus use a COMSOL model to find the actual RC contribution to our device's frequency response, plotted in the inset of Fig. \ref{freqresp}a. We also plot the product of the RC-limited frequency response and the $\tau_\mathrm{eph}$-limited frequency response with an assumed $(1+(2\pi\tau_\mathrm{eph} f)^2)^{-0.5}$ dependence, which applies as the electron-phonon cooling length $\ell=\sqrt{\kappa\tau_\mathrm{eph}/C_\mathrm{el}}\approx230\,\nm$ is narrower than our device channel\cite{jsong}. We thus predict a 3-dB cutoff frequency of $f_{-3\mathrm{dB}}\approx1.3\,\mathrm{GHz}$, dominated by the capacitance between the graphene back-gates.

To investigate our device's noise performance, we modulate the QCL current at $30\,\mathrm{kHz}$, amplify the photovoltage with a low-noise preamplifier and inspect using a signal analyzer. We observe in Device A no broadening of the $30\,\mathrm{kHz}$ photoresponse peak at offset frequencies as low as $0.1\,\mathrm{Hz}$, indicating long-term responsivity stability, and we observe no illumination-dependence of the noise floor. We then measure the un-illuminated noise spectral density and resistance versus both gate voltages. Fig. \ref{freqresp}b shows the resulting data for a Device B of identical design to Device A, organized by resistance and compared to the expected Johnson noise spectral density. We observe excellent consistency between the measured and predicted noise, with a $2\text{--}4\,\mathrm{dB}$ discrepancy consistent with the specified noise figure of our preamplifier, corroborating our earlier claim of Johnson-noise-limited NEP.

To demonstrate our device's utility, we analyze its predicted gas-sensing performance, summarized from Supplementary Section F. The minimum detectable gas concentration for a given waveguide platform and photodetector is given by\cite{sensoranalysis}:
\begin{equation}
p_\mathrm{gas,min}=e\,\frac{\alpha_\mathrm{base}\,\mathrm{NEP}}{an_\mathrm{g}\Gamma_\mathrm{E}I_0},
\end{equation}
where $I_0$ is the source power, $\alpha_\mathrm{base}$ is the waveguide attenuation coefficient in the absence of gas, $a$ is the specific attenuation coefficient of the gas, $n_\mathrm{g}$ is the guided mode group index, $\Gamma_\mathrm{E}$ is the confinement factor of electric field energy within the gaseous medium, and $e=\exp(1)$. For detection of nitric oxide (NO), with an absorption peak at $\lambda=5.24\,\um$ and a specific attenuation of approximately $a\approx70\,\mathrm{m}^{-1}\mathrm{atm}^{-1}$ at low concentrations\cite{nitricoxide}, we arrive at $p_\mathrm{gas,min}=74\,\textrm{\textmu atm}/\sqrt{\mathrm{Hz}}$ for a $1\,\mathrm{mW}$ illumination source. Assuming a measurement bandwidth of $0.1\,\mathrm{Hz}$ over which we have measured our photoresponse to be stable, we find $p_\mathrm{gas,min}=23\,\mathrm{ppm}$, roughly equal to the National Institute of Occupational Safety and Health (NIOSH) recommended exposure limit (REL) of $25\,\mathrm{ppm}$\cite{niosh}. Removing the slightly lossy HfO$_2$ dielectric underneath the gas-light interaction waveguide could decrease $p_\mathrm{gas,min}$ considerably, as waveguide losses down to $0.7\,\mathrm{dB}/\mathrm{cm}$ have been demonstrated at the same wavelength using a similar chalcogenide glass and liftoff process\cite{hongtaoliftoff}.

\begin{table}[th]
    \centering
    \begin{tabular}{c|cccc}
     & \twoline{HgCdTe PD}{$\lambda_\mathrm{opt}\!=\!5.0\,\um$} & \twoline{HgCdTe PD}{$\lambda_\mathrm{opt}\!=\!10.6\,\um$} & \twoline{VO$_\mathrm{x}$}{bolometer} & \twoline{This}{work} \\\hline
    \rule{0pt}{18pt}\twoline{NEP}{$[\mathrm{pW}/\sqrt{\mathrm{Hz}}]$} & \twoline{1, $\lambda\!=\!5.2\,\um$}{0.2, $\lambda\!=\!5.0\,\um$} & \twoline{10, $\lambda\!=\!5.2\,\um$}{40, $\lambda\!=\!10.6\,\um$} & 0.9 & 1100 \\
    \rule{0pt}{18pt}\twoline{$f_{-3\mathrm{dB}}$}{$[$MHz$]$} & 1.3 & 106 & $10\,\mathrm{Hz}$ & \twoline{1300}{(pred.)} \\
    \rule{0pt}{18pt}\twoline{Vacuum}{required?} & No & No & Yes & No \\
    \rule{0pt}{18pt}\twoline{Waveguide-}{integrated?} & No & No & No & Yes \\
    
    \end{tabular}
    \caption{Comparison of our detector with inferred room-temperature performance metrics for two HgCdTe photodiodes optimized for two different wavelengths (from \cite{rthgcdte}) and a VO$_x$ bolometer (from \cite{drsvox}) available off the shelf. For the photodiodes, the NEP is extrapolated from the specified detectivity for a detector scaled down to match the size of a diffraction-limited spot with $\mathrm{NA}=0.3$, which is the acceptance NA of these detectors. For the bolometer, we give the NEP of a single $17\,\um\times17\,\um$ bolometer pixel as calculated from the specified noise-equivalent temperature difference as described in Rogalski\cite{rogalski}.}
    \label{perfcomparison}
\end{table}

Although our demonstration is limited to $\lambda=5.2\,\um$ by light source availability, the optical conductivity of our graphene inferred from the fitting parameters in Table \ref{params} remains relatively constant and even increases at longer wavelengths due to intraband absorption as shown in Supplementary Fig. 7. We thus expect our platform to scale to $\lambda=10\,\um$ and beyond, perhaps requiring a BaF$_2$ substrate for extended transparency, with little reduction in performance owing to the PTE effect's thermal nature. In Table \ref{perfcomparison} we compare our device's performance with various off-the-shelf detectors. Although its NEP is not yet on par with commercial options, its predicted bandwidth may be useful for dual-comb spectroscopy-based integrated gas analyzers\cite{dcs}. Additionally, the vacuum requirement of VO$_x$ bolometers may complicate co-packaging and introduce coupling losses, and the high cost of HgCdTe may preclude use in broadly deployed sensor networks.

In conclusion, we have demonstrated a PTE-based graphene photodetector, integrated in a scalable chalcogenide glass waveguide platform with an NEP of $1.1\,\mathrm{nW}/\mathrm{Hz}^{1/2}$ and a bandwidth exceeding $f_{-3\mathrm{dB}}=1\,\mathrm{MHz}$. We have modeled the bandwidth to approach $1.3\,\mathrm{GHz}$ and we predict similar performance at longer wavelengths for scaled-up devices enabled by the transparency of GSSe beyond $\lambda=10\,\um$\cite{gesbse}. Finally, we have shown that our device and waveguide platform would enable NO detection at concentrations comparable to its REL. Substantial improvements are likely using metal-insulator-metal\cite{plasmonic} or dielectric slot waveguides to concentrate the optical mode to within a cooling length of the \emph{pn}-junction, which would also reduce the detector length and thus device footprint needed to absorb an optical signal. Gapped bilayer graphene may also be investigated as an alternative to monolayer graphene to reduce thermal noise\cite{bilayergr}. %We recommend future efforts to investigate shrinking the optical mode and graphene channel to the cooling length $\ell$ using metal-insulator-metal plasmonic waveguides\cite{plasmonic}, which could improve the NEP by constraining optical absorption to within..[EDIT, SEE COMMENTS] %increasing the utilization efficiency of absorbed light without imposing unrealistic material quality requirements \cite{jsong}.
%the following were added by Dirk:
The PIC platform further promises to support a full toolkit of mid-IR active devices including on-chip quantum cascade light sources\cite{qclonchip}, and may even leverage the same graphene material platform for devices such as graphene modulators\cite{hongtao} and hot-electron-based\cite{kim2021mid} or gapped bilayer graphene light sources. The platform could also be adapted to alternative mid-IR waveguide platforms, such as suspended Ge, as necessary to reach longer wavelength ranges\cite{suspendedge}.
% delete/clean up/below, we rewrote it above
%We also advocate further investigation of integrated MIR light sources to enable fully integrated systems. Sources based on super-Planckian plasmon-mediated emission in graphene\cite{kim2021mid} would be ideal, eliminating the requirement for heterointegration of III-V-based sources.
This research represents the first foray into waveguide-integrated detectors operating beyond $\lambda=4\,\um$, paving the way towards 2D-material-enabled integrated mid-IR microsystems for gas sensing, spectroscopy\cite{dcs} and free-space optical communications\cite{qclfso2}.
% By demonstrating an architecture that shows operation at 5 um and is readily extensibly beyond 10 um, we anti

\section*{Methods}

\subsection*{Photodetector Fabrication}

A continuous monolayer graphene film was grown on Cu foil (99.8\%, Alfa Aesar, annealed, uncoated, item no. 46365) cut to a size of $15\times2$ cm$^2$ in a 1-inch-diameter quartz tube furnace under atmospheric pressure. The furnace was heated to $1060^\circ$C over 30 minutes under 500 sccm of Ar flow; afterwards, 15 sccm of H$_2$ and 10 sccm of dilute CH$_4$ (1\% in Ar) were introduced as reducing gas and carbon source, respectively, and flowed for 4 hours to ensure the continuity of the graphene film. Finally, the furnace was allowed to cool to $100^\circ$C without modifying the gas flow before the CVD graphene was removed from the chamber. Our devices were fabricated on a 1" diameter by 1.0 mm thick (111)-cut CaF$_2$ substrate (MTI Corporation, item CFc25D10C2). We first coated our substrate with a PMMA bilayer for liftoff (495 PMMA A6 followed by 950 PMMA A2), which features a slightly re-entrant sidewall profile after developing. We then performed e-beam lithography using an Elionix FLS-125 125 keV electron beam lithography system to pattern alignment marks on our substrate, followed by room-temperature development in 3:1 isopropanol:methyl isobutyl ketone, e-beam evaporation of 5 nm Ti/100 nm Au (Temescal VES2550), and liftoff. To transfer the first layer of graphene, we first coated one side of the CVD graphene-on-Cu sheet with PMMA and removed the graphene from the other side using 90 seconds of oxygen RIE (16 sccm He and 8 sccm O$_2$ at a pressure of 10 mTorr and an RF power of $100$W, ``oxygen RIE process''). We then etched away the Cu using a FeCl$_3$-based etchant, followed by 2 DI water rinses, a 30-minute clean in 5:1 DI water:HCl 37\% in water to reduce metal ion contamination, and two more DI water rinses. After letting the graphene film sit overnight in the final evaporating dish of water, we scooped it out with our CaF$_2$ substrate, blew N$_2$ on the film to eliminate most of the trapped water, and then baked the sample at $80^\circ$ for 30 minutes followed by $160^\circ$ for 2 hours. We then removed the PMMA from the graphene using acetone at room temperature, rinsed it in isopropanol and blew it dry, and baked the sample in N$_2$ for 1 hour to improve adhesion. To pattern the graphene back-gates, we spun on a layer of 950 PMMA A6, patterned the gates in the Elionix and developed as described previously, etched away the exposed graphene using ``oxygen RIE process'', and removed the PMMA as described previously. We then repeated the Ti/Au liftoff process described above to define the contacts to the graphene gates, after which we evaporated 1.5 nm Al (Temescal VES2550) as an ALD seed layer, allowed the thin Al layer to oxide in ambient, and deposited 300 cycles $\approx 30$ nm of HfO$_2$ ALD at $200^\circ$C (Cambridge Nanotech Savannah 200). To define the graphene channel and channel contacts, we performed another graphene transfer as described above and repeated the subsequent graphene patterning and contact metallization steps, followed by another Al seed layer and 150 cycles of HfO$_2$ ALD to protect the graphene channel. Finally, to pattern the GSSe waveguides, we coated the chip with 495 PMMA A11, used the Elionix to define the waveguides, developed as described previously and evaporated 750 nm of Ge$_{28}$Sb$_{12}$Se$_{60}$, followed by a quick liftoff in boiling acetone, IPA rinse and N$_2$ blow-dry, and cleaving of the chip to expose waveguide facets.

\subsection*{Measurement conditions}

The maps in Figs. \ref{rawmeas}a, b, and c were measured by sequentially measuring each data point column by column, bottom to top from left to right. SR830 lock-in amplifiers were used for all measurements. Prior to each data point collection, both gate voltages were reset to $-7\,\mathrm{V}$ for $80\,\mathrm{ms}$ to reset the gate dielectric hysteresis (see Supplementary Section B), then set to the desired gate voltages and allowed to dwell for $200\,\mathrm{ms}$ for the lock-in signal to stabilize. The lock-in filter was set to a $30\,\mathrm{ms}$ time constant with a $12\,\mathrm{dB}/\mathrm{octave}$ falloff. The detector photovoltage in Fig. \ref{rawmeas}a was measured directly by the lock-in amplifier with no additional amplification. For the resistance map in Fig. \ref{rawmeas}b, we used our lock-in amplifier to bias the device with a $1\,\mathrm{VRMS}$ sine wave at $3.78\,\mathrm{kHz}$ through a $100\,\mathrm{k}\Omega$ resistor to act as a current source and measured the voltage across the device with the lock-in. To produce the frequency response plots in Fig. \ref{freqresp}a, we apply a sinusoid of variable frequency to the current modulation input of our QCL and measure the calibration and photoresponse signals with a SR844 RF lock-in amplifier. For the laser modulation response (indicated in red in Fig. \ref{freqresp}a), we couple the laser light through a single-mode waveguide on our chip with no devices on it and directly measure the amplified transmission signal produced by the fast InAsSb detector on the output side of our chip. For the photovoltage signal (blue curve in Fig. \ref{freqresp}a), we amplify the photovoltage produced by our detector by 40 dB using a preamplifier and measure this amplified signal with our lock-in. In all cases, we used a dwell time of $1.5\,\mathrm{s}$, and the filter of our lock-in was set to $100\,\mathrm{ms}$ with a $12\,\mathrm{dB}/\mathrm{octave}$ falloff. To measure the un-illuminated noise spectral density in Fig. \ref{freqresp}b, we amplify the noise produced by the device using a 60 dB preamplifier and analyze the output on an FFT signal analyzer while controlling the gate voltages applied to the device. We choose to measure the averaged noise spectral density between 22 and $32\,\mathrm{kHz}$ where we find no electromagnetic interference-related spectral peaks. We also use a lock-in amplifier to simultaneously measure the device resistance as described above for Fig. \ref{rawmeas}b, albeit at a higher frequency so as to not produce a signal in the noise measurement range. We use our signal analyzer's band averaging feature to measure the noise spectral density for each data point. To produce the final plot, we manually record the resistance and noise spectral density for all gate voltage pairs from $-6\,\mathrm{V}$ to $6\,\mathrm{V}$ in steps of $2\,\mathrm{V}$.

\subsection*{Device modelling}

We use the Kubo formula adapted from Hanson\cite{Hanson} to model graphene's conductivity at DC and infrared frequencies (albeit with different values of the Drude scattering time $\tau$ for the different frequency ranges):

\begin{multline}
\sigma(\omega,E_F,\tau,T)  = \frac{je^2(\omega-j\tau^{-1})}{\pi\hbar^2} \\ \times \Biggl[\frac{1}{(\omega-j\tau^{-1})^2}\int_0^\infty \varepsilon\left(\frac{\partial f_d(\varepsilon)}{\partial\varepsilon} - \frac{\partial f_d(-\varepsilon)}{\partial\varepsilon}\right) d\varepsilon \\ - \int_0^\infty \frac{f_d(-\varepsilon)-f_d(\varepsilon)}{(\omega-j\tau^{-1})^2-4(\varepsilon/\hbar)^2}d\varepsilon\Biggr]
\label{kubo}
\end{multline}
where $e$ is the elementary charge, $f_d(\varepsilon)=(\exp((\varepsilon-E_F)/k_BT) + 1)^{-1}$ is the Fermi-Dirac distribution and $k_B$ is Boltzmann's constant. As I will show below, graphene's low frequency conductivity $\sigma_\mathrm{DC}$ and infrared conductivity $\sigma_\mathrm{IR}$ affect various intermediate model parameters; $\sigma_\mathrm{DC}$ and $\sigma_\mathrm{IR}$ themselves depend strongly on $E_F$, which features spatial variation due to the back-gates. For the graphene channel, we assume a constant $N_\mathrm{c} = N_{0,\mathrm{c}}\,+\, e^{-1} C_\mathrm{g} V_\mathrm{g}$ in the region above each gate, where $N_\mathrm{c}$ is the carrier concentration in the channel (positive for positive $E_F$, negative for negative $E_F$), $N_{0,\mathrm{c}}$ is the native carrier concentration at zero gate voltage, $C_\mathrm{g}$ is the capacitance per area of the gate dielectric, and $V_\mathrm{g}$ is the voltage applied to the gate in question. (Using a set of test devices, we measure $C_\mathrm{g}=34.\,\mathrm{fF}/\um^2$ on our chip, corresponding to a back-gate dielectric constant of $K\approx12$; this is described in more depth in Supplementary Section D.) In the part of the graphene channel above the gap between the two gates, we assume a linear slope between $N_{\mathrm{c},1}$ and $N_{\mathrm{c},2}$. For the gates, $N_\mathrm{g} = N_{0,\mathrm{g}}\,-\,e^{-1} C_\mathrm{g} V_\mathrm{g}$, with $N_\mathrm{g}$ and $N_{0,\mathrm{g}}$ defined similarly to $N_\mathrm{c}$ and $N_{0,\mathrm{c}}$. In general, the graphene's Fermi level and carrier concentration are related by $E_F = \hbar v_\mathrm{gr} \sqrt{\pi |N|}\,\mathrm{sign}(N)$, where $v_\mathrm{gr}$ is graphene's Fermi velocity. To incorporate the blurring of the graphene's Fermi level-dependent properties due to spatial carrier concentration variations, we convolve the Kubo formula with a Gaussian as follows:
\begin{multline}
\sigma_\mathrm{DC}(N) = \\ \frac{1}{\sigma_\mathrm{n}\sqrt{2\pi}}\int_{-\infty}^\infty e^{-\frac{1}{2}\frac{(n-N)^2}{\sigma_\mathrm{n}^2}} \sigma(0,E_F(N),\tau_{DC},T_0) \,dn
\label{gaussian}
\end{multline}
and similarly for $\sigma_\mathrm{IR}(N)$ using $\omega=2\pi c/\lambda$ instead of 0 and $\tau_\mathrm{IR}$ instead of $\tau_\mathrm{DC}$. Finally, we have $R=\sigma_\mathrm{DC}^{-1}$, $\kappa=\pi^2 k_B^2 T_0 \sigma_\mathrm{DC}/3e^2$ via the Wiedemann-Franz law, and $S=-d(\log\sigma_\mathrm{DC})/dE_F$ \cite{grosso}. $C_\mathrm{el}$ is obtained by convolving the heat capacity of pristine graphene with a Gaussian of standard deviation $\sigma_N$ as in Eqn. \ref{gaussian}, where the pristine heat capacity is given by\cite{grosso,castroneto}:
\begin{equation}
C_\mathrm{el}(N)\Bigr|_{\sigma_\mathrm{n}=0} = \int_{-\infty}^{\infty} \varepsilon\, \frac{2|\varepsilon|}{\pi\hbar^2 v_\mathrm{gr}^2} \frac{\partial f_d(\varepsilon - E_F(N))}{\partial T}\,d\varepsilon.
\end{equation}

We use a waveguide eigenmode solver to find the mode profile of our waveguide at $\lambda = 5.2\,\um$, using refractive indices of 1.4, 2.6, and 1.88 for the CaF$_2$, GSSe, and HfO$_2$, respectively. The resulting mode profile enters into our expression for $\dot{Q}_\mathrm{el}$ as follows\cite{snyderandlove}:
\begin{equation}
\dot{Q}_\mathrm{el}=P\,\frac{\left(|E_x(x,y_\mathrm{c})|^2+|E_y(x,y_\mathrm{c})|^2\right) \sigma_\mathrm{IR,c}(x)}{\iint_{\mathbb{R}^2} \mathrm{Re}(\mathbf{E}\times\mathbf{H}^*)\cdot\hat{\mathbf{z}}\,dx\,dy}.
\end{equation}
Here $y_\mathrm{c}$ is the $y$-coordinate of the graphene channel, and $y_\mathrm{g}$ would be the $y$-coordinate of the graphene gates. We may then write $\alpha_\mathrm{c} = P^{-1}\,\int_{-W/2}^{W/2} \dot{Q}_\mathrm{el}(x)\,dx$. Similar expressions hold for $\alpha_\mathrm{g}$ in terms of $\sigma_\mathrm{IR,g}(x)$, noting of course that $\sigma_\mathrm{IR,g}(x)=0$ for $x$ within the gap between the gates where there is no graphene. Finally, $\rho_\Omega=\int_{-W/2}^{W/2} R(x)\,dx$.

Having thus obtained expressions for $\kappa(x)$, $C_\mathrm{el}(x)$, $\dot{Q}_\mathrm{el}(x)$, $S(x)$, $\Pi(x)$, $\alpha_\mathrm{c}$, $\alpha_\mathrm{g}$ and $\rho_\Omega$ as a function of the gate voltages as well as $\tau_\mathrm{DC}$, $\tau_\mathrm{IR}$, $\sigma_\mathrm{n}$, $E_\mathrm{Fc}$, $E_\mathrm{Fg}$, $\tau_\mathrm{eph}$, $\alpha_\mathrm{e}$, and $\rho_\mathrm{c}$, we then solve for the increase in electronic temperature per guided power $\Delta T_\mathrm{el}(x)/P = (T_\mathrm{el}(x) - T_0)/P$ using the equation:

\begin{equation}
-\frac{d}{dx}\left(\kappa\,\frac{d\,\Delta T_\mathrm{el}}{dx}\right) + \tau_\mathrm{eph}^{-1} C_\mathrm{el} \Delta T_\mathrm{el} = \eta\, \dot{Q}_\mathrm{el} - J_x\frac{d\Pi}{dx},
\label{ode}
\end{equation}

where $\kappa$ is the 2D electronic thermal conductivity of the graphene, $\tau_\mathrm{eph}$ is the electron-phonon cooling time, $\dot{Q}_\mathrm{el}$ is the absorbed optical power per area, $\eta$ is the conversion efficiency of absorbed optical power to electronic heat after initial electron-phonon scattering\cite{jsong}, $J_x$ is the line current density in the $x$-direction, and $\Pi$ is the Peltier coefficient. We are approximating the electric field to run exclusively in the $x$-direction, valid for sufficiently gradual light absorption. We assume $\eta=1$ as has been previously reported in pump-probe experiments at this wavelength range\cite{carriercarrier}. The thermal electromotive force (EMF) arising from the Seebeck effect is then given by:

\begin{equation}
\mathscr{E}_x = -\int_{-W/2}^{W/2} S \, \frac{d\,\Delta T_\mathrm{el}}{dx}\,dx,
\label{emf}
\end{equation}

where $W=5.4\,\um$ is the channel width and $S$ is the Seebeck coefficient. In Eqns. \ref{ode} and \ref{emf}, $\kappa$, $C_{el}$, $S$, and $\Pi = ST_\mathrm{el} \approx ST_0$ (for small $\Delta T_\mathrm{el}$) are all dependent on the local Fermi level $E_\mathrm{F}$ of the graphene, and thus have a gate-tunable $x$-dependence, which we account for in our calculations. Combining the equations, the $\eta \dot{Q}_\mathrm{el}$ source term in Eqn. \ref{ode} gives rise to a proportional photo-induced EMF, whereas the Peltier term $J_x\frac{d\Pi}{dx}$ gives rises to a current-dependent EMF which appears as a resistance in series with the Ohmic and contact resistances of the channel. We can thus write:

\begin{equation}
V = \overline{\mathcal{R}}_v\,\alpha_\mathrm{c}\,P(z) - \left(\rho_\Omega + \rho_\Pi + \rho_\mathrm{c}\right) J_x(z)
\label{emf2}
\end{equation}

where $V$ is the voltage across the contacts, $\overline{\mathcal{R}}_v$ is the photovoltage per absorbed power per length of a cross-sectional slice of the device (i.e., dimensions of V/(W/m)), $\alpha_\mathrm{c}$ is the component of the waveguide power attenuation coefficient arising from absorption in the graphene channel, $P(z)$ is the guided power at a position along the waveguide, and $\rho_\Omega$, $\rho_\Pi$, $\rho_\mathrm{c}$ are the Ohmic, Peltier, and contact line resistivities (dimensions of $\Omega\cdot\mathrm{m}$), respectively. Averaging over $z$ along the length of the waveguide we obtain:

\begin{equation}
V = \frac{\overline{\mathcal{R}}_v\,\alpha_\mathrm{c}}{L\,\alpha_\mathrm{tot}}\left(1-e^{-\alpha_\mathrm{tot}L}\right)P_\mathrm{in} - \left(R_\Omega + R_\Pi + R_\mathrm{c}\right) I,
\label{emf3}
\end{equation}

where $I$ is the current produced by the photodetector, thus describing a Th\'evenin equivalent source. Here, $\alpha_\mathrm{tot}=\alpha_\mathrm{c}+\alpha_\mathrm{g}+\alpha_\mathrm{e}$ is the total guided power attenuation coefficient within the detector, including contributions not only from the graphene channel but also from the graphene gates ($\alpha_\mathrm{g}$) as well as a gate-independent excess loss $\alpha_\mathrm{e}$ associated with scattering and absorption from organic or metallic impurities attached to or trapped underneath the graphene sheets. Thus the total device resistance is equal to $R=R_\Omega + R_\Pi + R_\mathrm{c}$, and the voltage responsivity is given by:

\begin{equation}
\mathcal{R}_v = \frac{\overline{\mathcal{R}}_v\,\alpha_\mathrm{c}}{L\,\alpha_\mathrm{tot}}\left(1-e^{-\alpha_\mathrm{tot}L}\right),
\label{responsivity}
\end{equation}

%using Eqns. \ref{ode}--\ref{emf3}, as originally explained in Section \ref{model}. 
which we plot versus both gate voltages in Fig. \ref{comparison}b for the best-fit device parameters given in Table \ref{params} obtained as described in Supplementary Section C. All calculations are carried out in Mathematica. 
\section*{Author Contributions}

J.H., D.E. and J.G. conceived the experiments. J.G. designed, fabricated, and measured the devices, with the exception of chalcogenide glass deposition, performed by H.L. and S.D.-J. under the supervision of J.H., and graphene growth, performed by M.H. and A.-Y.L. under the supervision of J.K. and T.P. K.R. provided the chalcogenide glass sources for thermal deposition. J.G. and D.E. wrote the manuscript. All work was supervised by D.E.
\section*{Data Availability}

The datasets generated during and/or analysed during the current study are available in the FigShare repository at \url{https://doi.org/10.6084/m9.figshare.c.5514759.v1}.
\section*{Code Availability}

The Mathematica document used to simulate photodetector performance metrics is available in the FigShare repository at \url{https://doi.org/10.6084/m9.figshare.c.5514759.v1}.

\begin{acknowledgments}
We would like to thank the MIT.Nano and MIT Nanostructures Laboratory staff for maintaining the cleanroom facilities used to fabricate these devices, in particular Mark Mondol, Jim Daley, and Dave Terry. We also would like to thank Sebasti\'an Castilla of ICFO for helpful discussions. This research was funded in part by a grant from the Army Research Office via the MIT Institute for Soldier Nanotechnologies University-Affiliated Research Center (ISN UARC) (award number W911NF-18-2-004), the NSF Graduate Research Fellowship Program (award number 1122374), and NSF Award \#2023987. Any opinions, findings, and conclusions or recommendations expressed in this material are those of the author(s) and do not necessarily reflect the views of the National Science Foundation.
\end{acknowledgments}

\bibliography{wimp}% Produces the bibliography via BibTeX.

\end{document}

\maketitle

\section{Optical Setup, Alignment, and Power and Loss Calibration}

\begin{figure}[bh]
    \centering
    \includegraphics[width=\textwidth]{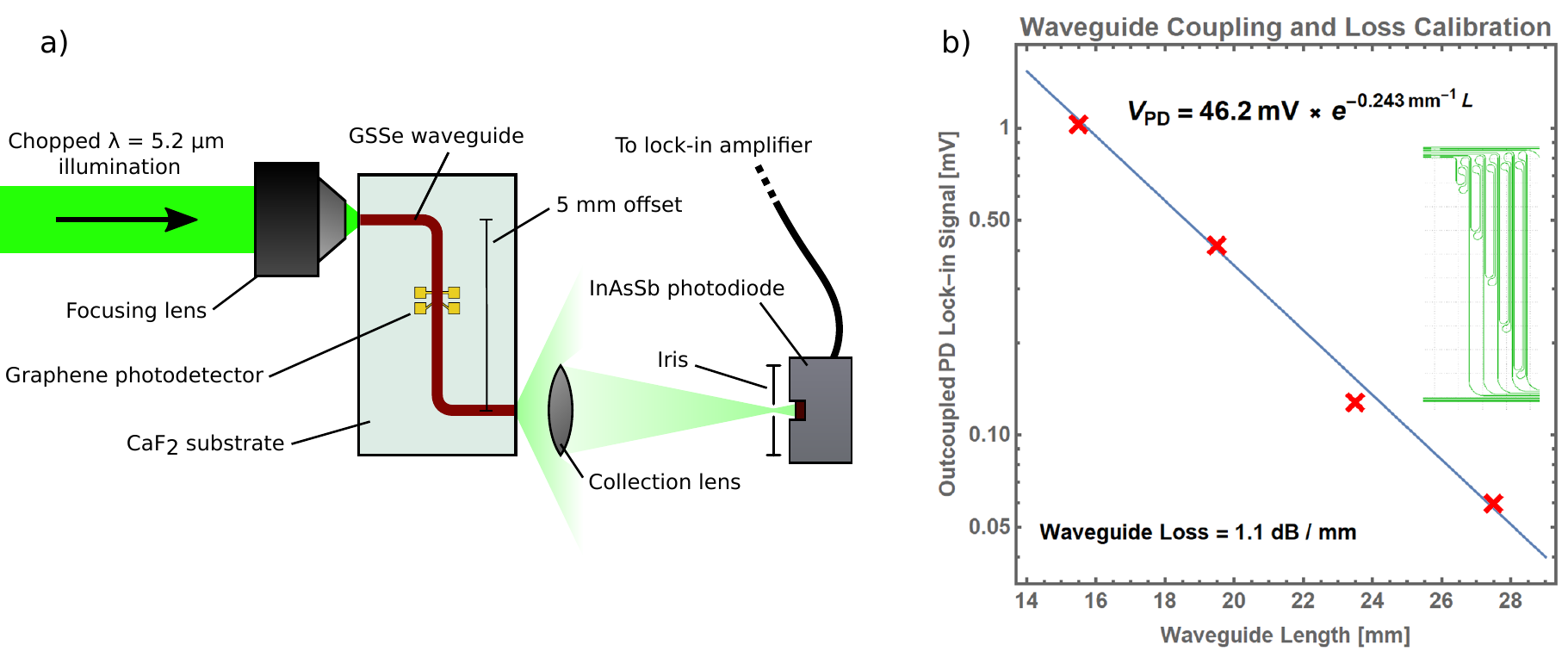}
    \caption{a) Depiction of the in- and out-coupling beam path for optical and optoelectronic measurements. Lengths are not to scale; the chip and waveguide are magnified for clarity. b) Output photodiode signal versus waveguide length for four waveguide kickback structures. The decaying exponential fit reveals the waveguide loss and the in-coupled power, subject to a conversion efficiency accounting for the collection efficiency of the collection optics and the voltage responsivity of the photodiode. Inset: Layout of the kickback structures.} %The longest of the five kickbacks was not included in the fit as the out-coupled signal was indistinguishable from stray light.}
    \label{setuppic}
\end{figure}

Supplementary Fig. \ref{setuppic}a depicts the optical beam path during all optical and optoelectronic measurements. Chopped, collimated illumination at a wavelength of $\lambda=5.2\,\um$ is coupled into a cleaved waveguide facet at the edge of the chip using a molded aspheric focusing lens to achieve a diffraction-limited spot. The waveguides containing our photodetectors and test devices as well as the kickback waveguides for loss measurement are designed to exit the chip at a 5 mm offset with respect to the input to reduce the amount of stray light picked up by the collection optics. The collection optics consists of a 1" high-NA germanium asphere which images the output facet at the center of an iris, behind which we place an InAsSb photodiode that monitors the out-coupled power. To align the setup, we flip the Ge collection lens, iris and photodiode out of output the beam path and use a long focal length CaF$_2$ lens and liquid nitrogen-cooled InAsSb camera to image the output facet of the chip while adjusting the chip position to achieve coupling first through a straight multimode waveguide (not shown) followed the desired S-shaped device or kickback waveguide. We then replace the long focal length lens with the high-NA germanium collection lens, flip the iris back into the collection beam path in the ``open'' position, and place the camera behind the iris. By adjusting the position of the collection lens to focus the out-coupled light at the center of the iris while gradually reducing the iris aperture size and monitoring the focal point on the camera, we are able to localize the focus of the collection lens at the center of the iris. Finally, we replace the camera with the InAsSb photodiode and adjust the in-coupling and out-coupling optics to maximize the signal measured by the photodiode.

\begin{figure}[b!]
    \centering
    \includegraphics[width=\textwidth]{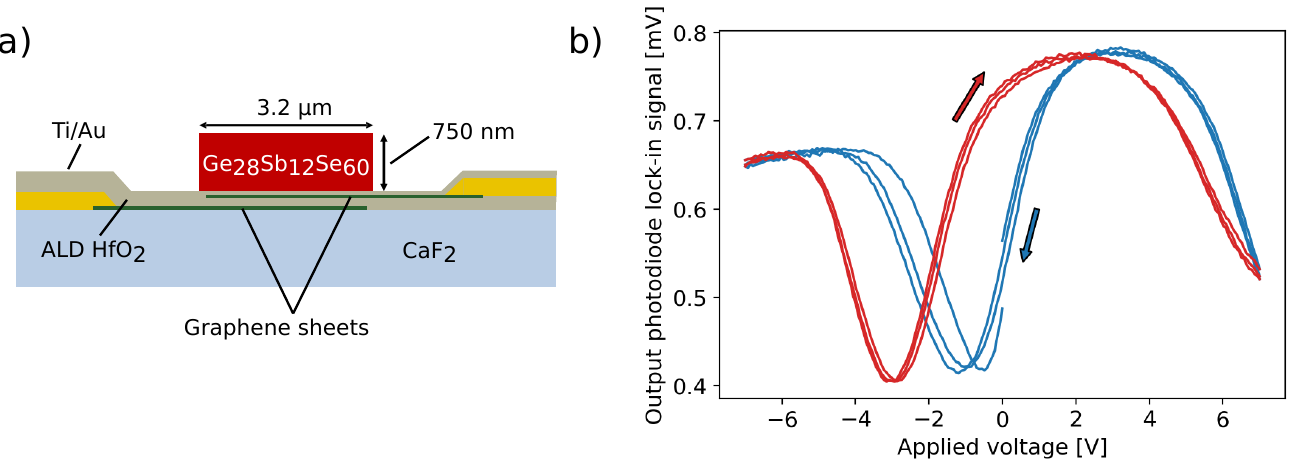}
    \caption{a) Illustration of the test device cross-section perpendicular to the waveguide axis. b) Transmission versus applied voltage for one of our test devices using a zigzag voltage sweep to reveal hysteresis. Sweep direction is color-coded and indicated with arrows.}
    \label{mod}
\end{figure}

To determine the loss of our waveguides and predict the optical power immediately incident upon our graphene photodetector during characterization, we have fabricated a set of five waveguide ``kickback'' structures of varying length, the layout of which is shown in the inset of Supplementary Fig. \ref{setuppic}b. A large radius of $75\,\um$ is used for the waveguide turns to reduce the associated light leakage to negligible levels. We individually align and optimize the coupling for each of these waveguides and record the lock-in signal of the collection photodiode; these data are plotted versus the total kickback waveguide length in Supplementary Fig. \ref{setuppic}b. We could not collect a data point for the longest kickback because we found the out-coupled light to be indistinguishable from stray illumination. Fitting the data to a decaying exponential reveals a waveguide loss of $1.1\,\mathrm{dB}/\mathrm{mm}$ and an in-coupled power corresponding to a lock-in signal of $46.2\,\mathrm{mV}$. To obtain the actual in-coupled optical power, we must divide this value by the product of the voltage responsivity of the photodiode and the efficiency of the collection optics. For the former, we obtain a value of $1.76\times10^3\,\mathrm{V}/\mathrm{W}$ from the manufacturer-provided calibration data. For the latter, we perform a full-wave electromagnetic simulation of the waveguide facet and extract the far-field profile of out-coupled light. Integrating the simulated optical power falling within the entrance pupil of the collection optics, we obtain a collection efficiency of $0.45$. Combining the above three figures, we arrive at $58.\,\mathrm{mW}$ coupled into the waveguide facet at optimal alignment. Accounting for waveguide loss, we obtain a power of $11.\,\mathrm{mW}$ immediately incident upon our measured photodetector, from which we can thus calculate the responsivity figures reported in the main article.

\section{Hysteresis effect}\label{hyst}

In addition to photodetector devices, we also fabricated gated graphene test devices as illustrated in Supplementary Fig. \ref{mod}a. These are similar to the detector devices described in the main article, except with only a single graphene sheet and a single contact on each graphene layer. Indeed, the detector devices can be made to behave similarly by applying the same voltage to each of the back-gates, but we report the test devices to illustrate the gate hysteresis effect observed for all devices. Supplementary Fig. \ref{mod}b shows the transmission response versus applied voltage for one of our test devices. Applying a zigzag voltage sweep reveals an undesirable hysteresis pattern showing distinct curves for rising and falling voltages sweeps, which we label red and blue respectively. The same effect is observed in our photodetector devices as well. Hysteresis has been widely reported for as-deposited HfO$_2$ gates, and is attributed to trapped charge carriers within the dielectric\cite{hfo2, hik}. To prevent the effects of hysteresis from appearing in the gate sweeps presented in Main Figure 2, we ``reset'' both gate voltages to $6.5\,\mathrm{V}$ prior to collecting each data point.

\section{Measured versus modelled resistance and transmittance and parameter extraction}

\begin{figure}[tb]
    \centering
    \includegraphics[width=0.9\textwidth]{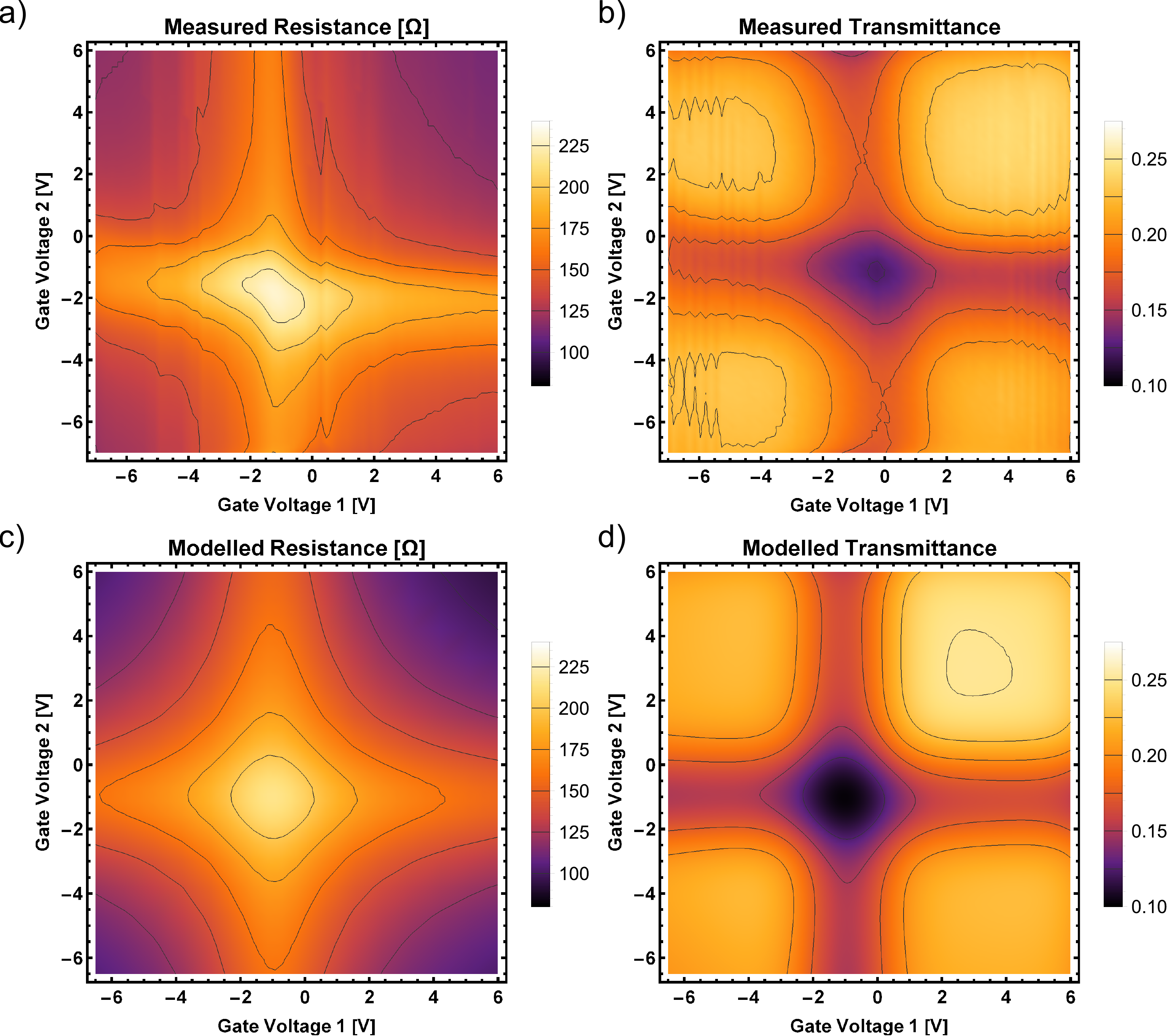}
    \caption{a, b) Contour plots of the measured resistance and transmittance maps as a function of gate voltages. c, d) Contour plots of the modelled resistance and transmittance maps as a function of gate voltages using the graphene quality and waveguide loss parameters listed in Main Table 1.}
    \label{comp}
\end{figure}

Supplementary Fig. \ref{comp} compares the measured and modelled resistance and transmittance gate maps obtained using the mean graphene quality and waveguide loss parameters listed in Main Table 1; namely, $\tau_\mathrm{DC}=3.5\,\mathrm{fs}$, $\tau_\mathrm{IR}=40\,\mathrm{fs}$, $\sigma_\mathrm{n}=2\times10^{12}\,\mathrm{cm}^{-2}$, and $\alpha_\mathrm{e}=2.5\,\mathrm{mm}^{-1}$. We determined the values of these parameters to achieve the best fit simultaneously between both pairs of maps. Generally speaking, $\tau_\mathrm{DC}$ is inversely proportional to $R_\Omega$, $\tau_\mathrm{IR}$ affects the scale and modulation contrast of the transmittance $\mathcal{T} = e^{-\alpha_\mathrm{tot}L}$, and $\sigma_N$ affects the sharpness (width at half-max) and contrast of both. First, we infer $E_\mathrm{Fc}$ simply from from the gate voltage of the charge neutral point (peak in the case of $R_\Omega$ and dip in the case of $T$). Since the sharpnesses of both $R$ and $\mathcal{T}$ are largely determined by $\sigma_N$ for relatively high $\sigma_N$ as is the case for our devices, we then determine $\sigma_N$ to best match both maps. In particular, we find that the resistance map is best fit by $\sigma_\mathrm{n}=1.5\times10^{12}\,\mathrm{cm}^{-2}$ and the transmittance map by $\sigma_\mathrm{n}=2.5\times10^{12}\,\mathrm{cm}^{-2}$, from which we obtain the error margins quoted for $\sigma_\mathrm{n}$ in Main Table 1; $\sigma_\mathrm{n}=2\times10^{12}\,\mathrm{cm}^{-2}$ represents a compromise between these two values. We then determine $\tau_\mathrm{DC}$ to roughly match the scale of $R$, but we allow the modelled resistance to be on the order of $10\,\Omega$ less the measured resistance to take into account the possibility of a contact resistance in this range as justified in Supplementary Section \ref{cres}. The uncertainty in the actual contact resistance, as well as the imperfect fit, both contribute to uncertainty in the actual value of $\tau_\mathrm{DC}$. We determine $\tau_\mathrm{IR}$ and $E_\mathrm{Fg}$ to best match the measured transmittance map, and the values quoted in Main Table 1 represent our best attempt to simultaneously reflect multiple features of this map; namely, the modulation contrast between the center and corners of the map, the contrast between the upper right and other corners, and the falloff at the low-voltage edges of the map (where intraband absorption of the graphene gates is strongest) and the high-voltage edges of the map (where interband absorption of the graphene gates is strongest). The quoted error margins of $\tau_\mathrm{IR}$ reflect the range over which these different features of the transmittance map are best rendered in our model, and the error margins of $\alpha_\mathrm{e}$ reflect the range required to match the scale of the measured transmittance map over the error range of $\tau_\mathrm{IR}$. We finally use these six parameters to predict the relative gate dependence of the voltage responsivity. In this way, only the overall scale factor of the responsivity is subject to a fitting parameter (namely, $\tau_\mathrm{eph}$); the contour of the responsivity map is, in our device's performance regime, purely predicted from parameters extracted from the resistance and transmittance maps. Therefore, the resemblance between the measured and modelled responsivity maps is not the result of tweaking parameters to achieve a fit, but rather reflects the accuracy of the photoresponse mechanism model itself using model parameters ``fed forward'' from the resistance and transmittance maps.

\section{Measurement of gate capacitance}

\begin{figure}[bt]
    \centering
    \includegraphics[width=0.5\textwidth]{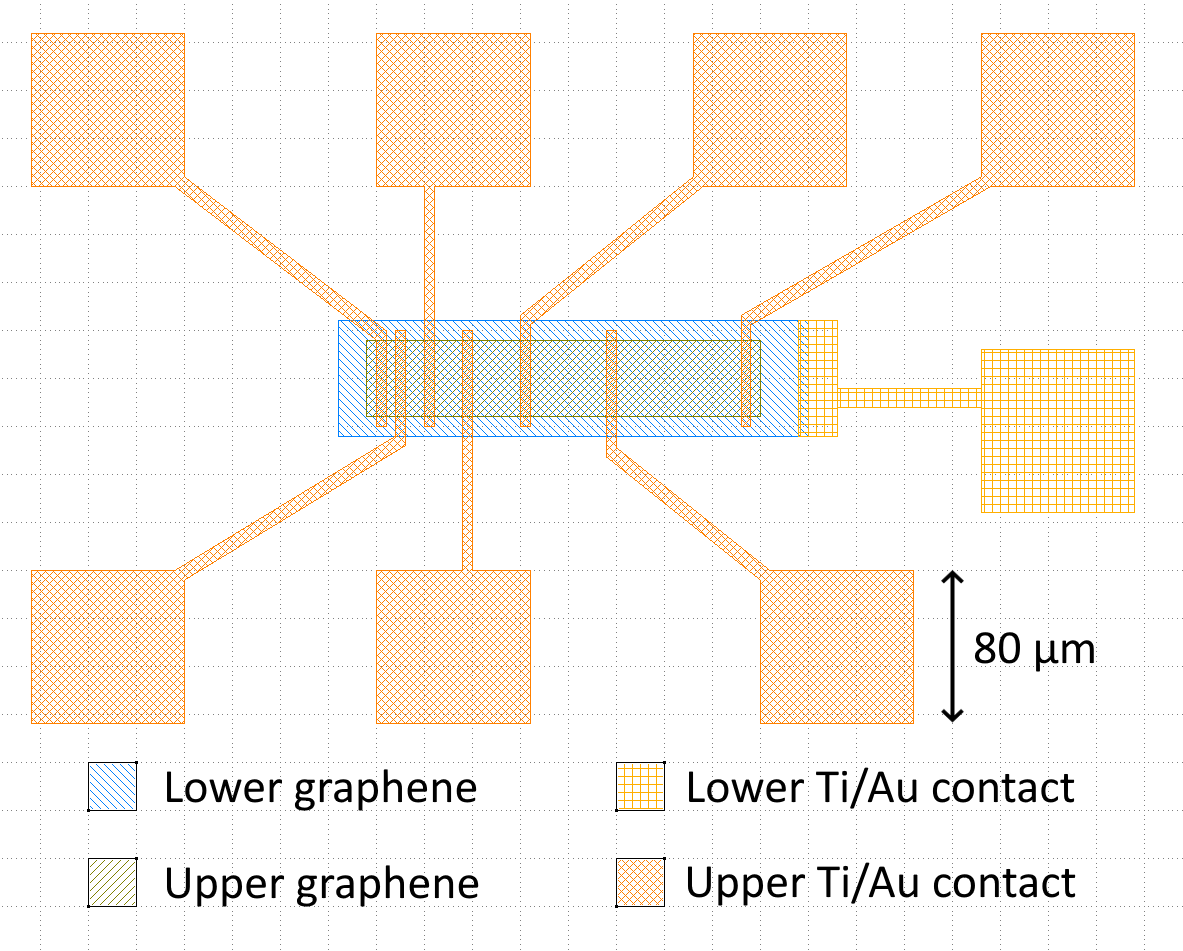}
    \caption{Transmission line method (TLM) device used for gate capacitance and contact resistance measurements.}
    \label{tlm}
\end{figure}

We used Transmission Line Method (TLM) devices to measure both the gate capacitance and contact resistance of our devices. Shown in Supplementary Fig. \ref{tlm}, these devices consist of a large, contacted graphene back gate, gating a set of graphene FETs of increasing channel lengths, ranging from $5\,\um$ to $65\,\um$. To extract the capacitance per area of our hafnia dielectric, we measure the capacitance between the lower graphene gate and the upper graphene/gold structure, and divide by the overlap area of these two structures. To measure the capacitance, we use a lock-in amplifier to apply a 0.5 VRMS sinusoid of variable frequency to the rightmost two metal contacts in Supplementary Fig. \ref{tlm}, and in the current return path we place a $10\,\textrm{k}\Omega$ shunt resistor, the voltage across which we monitor with the lock-in amplifier. At each frequency and for each TLM that we measure, we perform a measurement with both pads contacted and with only one pad contacted to compensate for any stray capacitance in our setup. We then subtract the measured capacitances in the ``connected'' and ``disconnected'' cases to obtain the actual device capacitance as a function of frequency, which we plot in Supplementary Fig. \ref{capacmeas}. Excluding the orange curve as an outlier, we measure a capacitance of $C\approx30\,\mathrm{pF}$ corresponding to a capacitance per area of $C_\mathrm{g}=3.4\,\mathrm{fF}/\um^2$ and a dielectric constant of $K\approx 12$.

\begin{figure}[bt]
    \centering
\includegraphics[width=0.7\textwidth]{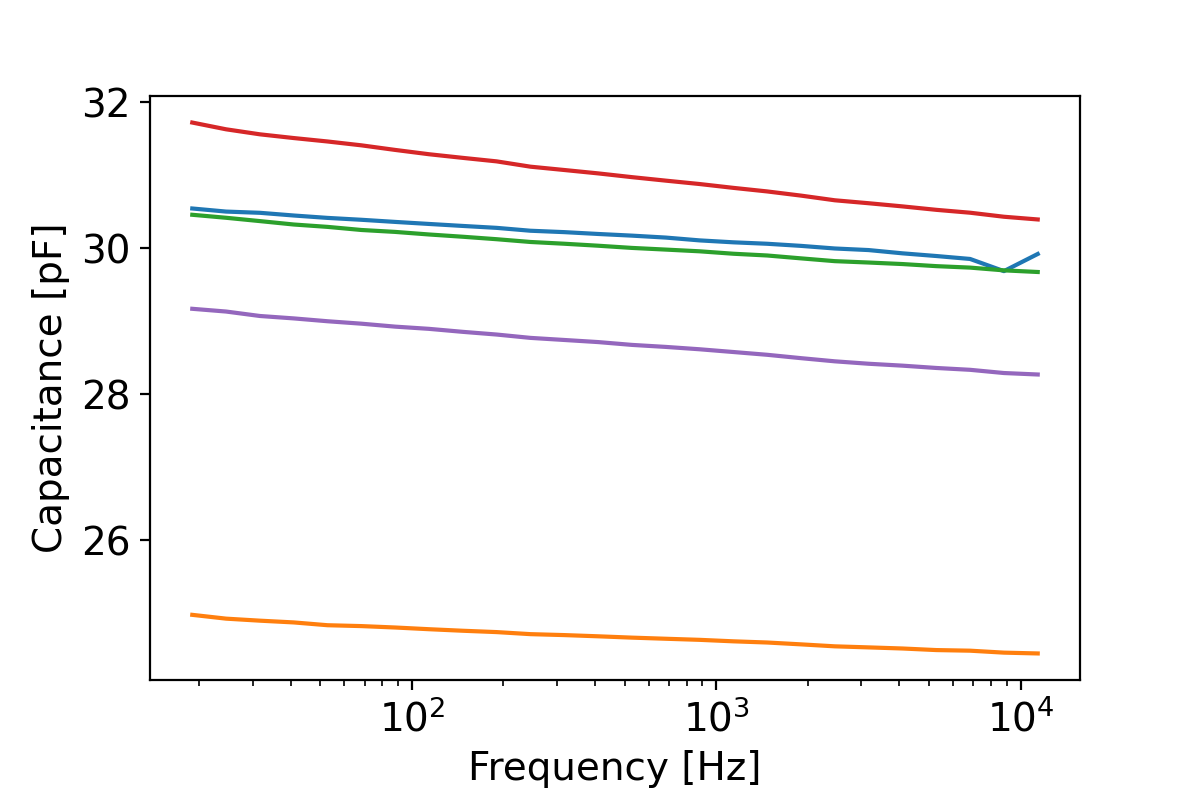}
 \caption{Capacitance versus measurement frequency for five devices.}
\label{capacmeas}
\end{figure}
% 
% \begin{figure}[h!]
%     \centering
%     \begin{circuitikz}
%     \draw
%     (0,0) to [vsourcesin, l^=$0.5\,\mathrm{V\,RMS}$] (0,3)
%     to (3, 3)
%     to [short, *-] (6,3)
%     to [generic, l^=$Z_\mathrm{DUT}$] (6,0)
%     to [short, -*] (3,0)
%     (0,0) to [R, l^=$R_\mathrm{shunt}$, v_=$V_\mathrm{meas}$] (3,0)
%     to [C, l_=$C_\mathrm{wiring}$] (3,3);
%     \end{circuitikz}
%     \caption{Circuit model depicting capacitance measurement.}
%     \label{cap}
% \end{figure}
% 
% Solving the circuit, we find:
% 
% \begin{equation}
% Z_\mathrm{DUT} = \left(\frac{1}{Z_\mathrm{con} - \textrm{Re}[Z_\mathrm{dis}]} - \frac{1}{\textrm{Im}[Z_\mathrm{dis}]}\right)^{-1}
% \label{zduteqn}
% \end{equation}
% 
% where $Z_\mathrm{con}$ is the impedance measured with both probes landed and $Z_\mathrm{dis}$ is the impedance measured with one probe slightly lifted off the probe pad. From the real part of $Z_\mathrm{DUT}$, we finally extract the inferred capacitance versus frequency for five such devices, the results of which are shown in Supplementary Fig. \ref{capacmeas}. At low frequencies, 
% 
% \begin{figure}[bt]
%     \centering
% includegraphics[width=0.7\textwidth]{capacmeas.png}
%     \caption{Capacitance extracted from Eqn. \ref{zduteqn} versus measurement frequency for five devices.}
%     \label{capacmeas}
% \end{figure}

\section{Measurement of TLM structure for gate resistance extraction}\label{cres}

In additional to capacitance measurement, we also use the TLM structures for their intended purpose of evaluating the resistance of our graphene-metal contacts. For each of the six channels in each of the five TLM devices, we measure the channel resistance as a function of gate voltage using an upward voltage sweep each time to compensate for the hysteresis discussed in Supplementary Section \ref{hyst}. Since the gate voltage of the charge neutral point may shift slightly between measurements due to trapped charges in the gate, we then shift the measured resistance curves so that their peaks overlap. Finally, for each measured voltage offset from the Dirac peak and for each TLM, we fit the data of resistance versus channel length to a line and plot the resulting y-intercept as a function of voltage offset, manually eliminating any gate voltage sweeps showing malformed (for instanced, flattened or bimodal) resistance peaks. The resulting y-intercept curves are shown in Supplementary Fig. \ref{tlmmeas}. Unfortunately, we find highly inconsistent intercept resistances between the five TLM devices, with the intercept even going negative in several cases. Therefore, we are unable to draw a quantitative conclusion regarding the contact resistance of our devices. We can at least, however, estimate the total contact resistance (summing over both contacts) for our TLM devices to be generally in the $\approx 1\times10^2\,\Omega$ range; therefore, since the TLM channels are $40\,\um$ wide, we would expect the total contact resistance of our actual photodetectors to be in the $\approx 1\times10^1\,\Omega$ range, which is an order of magnitude lower than our measured resistances; therefore we conclude that it can be safely ignored in our modelling as other sources of error (such as the imperfect fit between our measured and modelled resistance and transmittance maps) are much more likely to dominate the uncertainty in our analysis.

\begin{figure}[bt]
    \centering
    \includegraphics[width=0.7\textwidth]{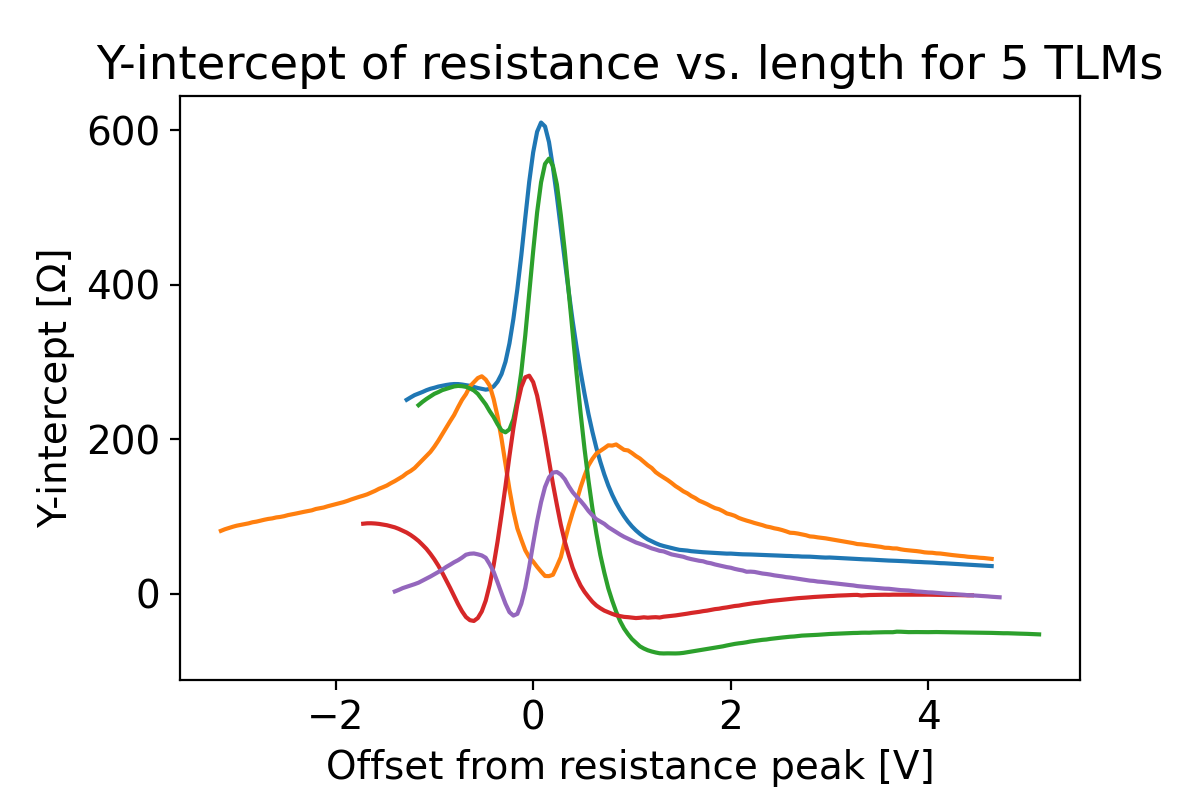}
    \caption{Y-intercepts of resistance versus channel length as a function of gate voltage offset from the resistance peak for our five TLM devices.}
    \label{tlmmeas}
\end{figure}

\section{Optical mode attenuation due to gas absorption and gas sensitivity analysis}

Our gas sensitivity analysis follows that of Siebert at el.\cite{sensoranalysis}, which considers a light source at the gas absorption peak illuminating a long gas-light interaction waveguide, partially cladded with the ambient air in which gas's presence is suspected, and terminated in a noisy photodetector. However, the authors do not give a rigorous analysis of the attenuation coefficient of guided light due to the absorbing gas, which we supply here. The power attenuation coefficient in a waveguide with a perturbative source of absorption is given by \cite{snyderandlove}:

\begin{figure}[tb]
    \centering
    \includegraphics[width=0.7\textwidth]{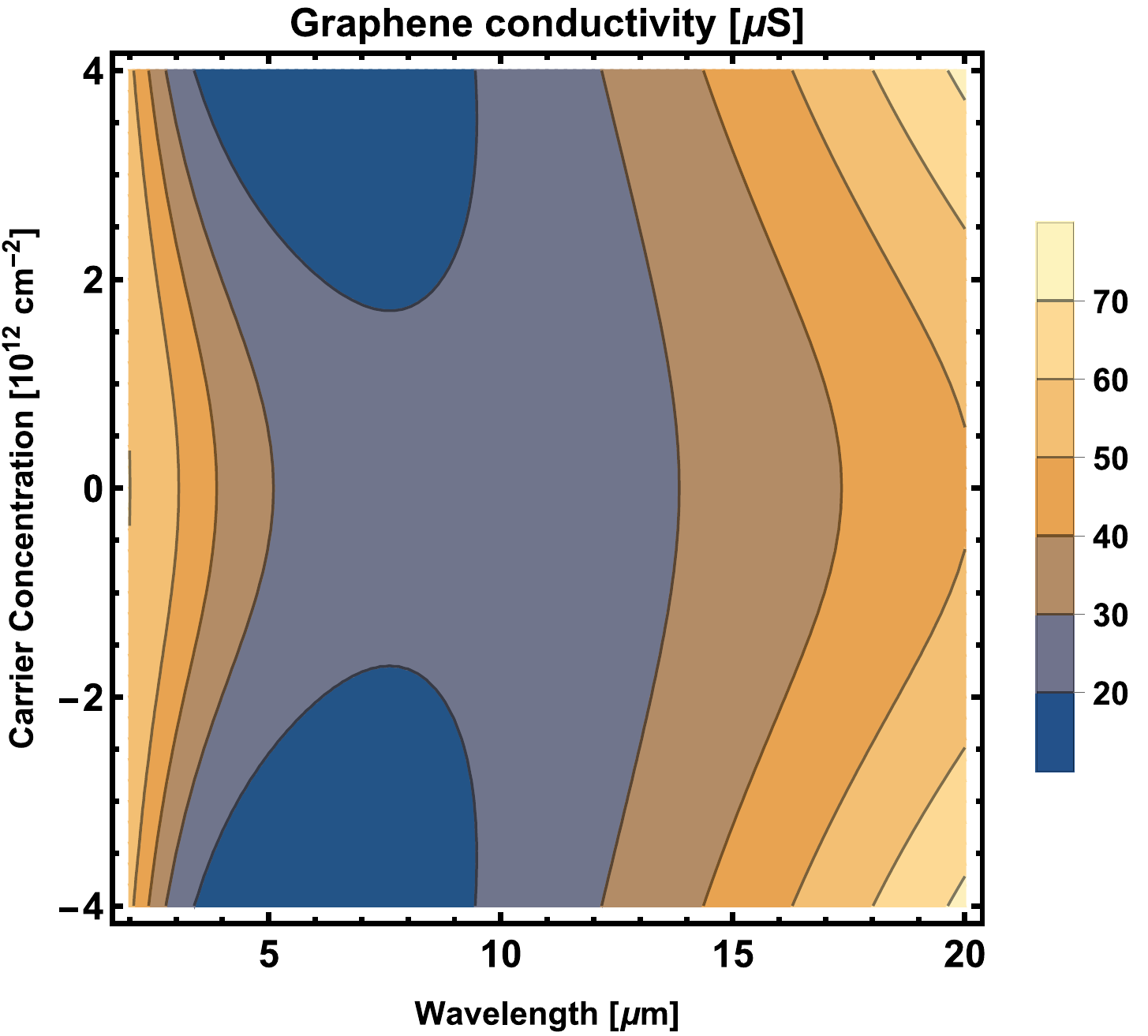}
    \caption{Contour plot of real part of graphene's infrared optical conductivity as a function wavelength and mean carrier concentration, assuming a carrier concentration spread of $\sigma_\mathrm{n}=2.0\times10^{12}\,\mathrm{cm}^{-2}$ and a Drude scattering time of $\tau_\mathrm{IR}=40\,\mathrm{fs}$.}
    \label{grcond}
\end{figure}
\begin{equation}
\alpha_\mathrm{wg} = \omega_0\,\frac{\iint_{\mathbb{R}^2}2n\kappa\,|\mathbf{E}|^2\,dx\,dy}{\iint_{\mathbb{R}^2} \mathrm{Re}(\mathbf{E}\times\mathbf{H}^*)\cdot\hat{\mathbf{z}}\,dx\,dy},
\end{equation}
where the complex refractive index $\bar{n}\equiv n-j\kappa$. For a partial pressure of target gas $p_\mathrm{gas}$, the free space attenuation coefficient is:
\begin{equation}
\alpha_\textrm{free-space} = ap_\mathrm{gas} = 2\kappa\omega_0/c,
\end{equation}
therefore $\kappa=ap_\mathrm{gas}c/(2\omega_0)$, and thus
\begin{equation}
\alpha_\mathrm{gas} = ap_\mathrm{gas}c\,\frac{\iint_\mathrm{gas}n_0\,|\mathbf{E}|^2\,dx\,dy}{\iint_{\mathbb{R}^2} \mathrm{Re}(\mathbf{E}\times\mathbf{H}^*)\cdot\hat{\mathbf{z}}\,dx\,dy},
\label{agas}
\end{equation}
where $n_0\approx1$ is the refractive index of the gaseous medium, as $\kappa$ is only nonzero in the gaseous region of the waveguide cross-section. Since our waveguide mode is absorbed only very weakly, we now reference the expressions for group velocity $v_\mathrm{g}$ and propagation constant $\beta$, respectively, given in Snyder and Love for non-absorbing waveguides\cite{snyderandlove}:
\begin{equation}
v_\mathrm{g}=\frac{c^2\beta}{\omega_0}\frac{\iint_{\mathbb{R}^2} \mathrm{Re}(\mathbf{E}\times\mathbf{H}^*)\cdot\hat{\mathbf{z}}\,dx\,dy}{\iint_{\mathbb{R}^2}n^2\,\mathrm{Re}(\mathbf{E}\times\mathbf{H}^*)\,dx\,dy},
\end{equation}
and
\begin{equation}
\beta = \frac{\omega_0}{c^2}\frac{\iint_{\mathbb{R}^2}n^2\,\mathrm{Re}(\mathbf{E}\times\mathbf{H}^*)\,dx\,dy}{\iint_{\mathbb{R}^2}n^2\,|\mathbf{E}|^2\,dx\,dy}.
\end{equation}
Combining these two equations, the group index $n_\mathrm{g}$ is thus:
\begin{equation}
n_\mathrm{g} = \frac{c}{v_\mathrm{g}} = c\,\frac{\iint_{\mathbb{R}^2}n^2\,|\mathbf{E}|^2\,dx\,dy}{\iint_{\mathbb{R}^2} \mathrm{Re}(\mathbf{E}\times\mathbf{H}^*)\cdot\hat{\mathbf{z}}\,dx\,dy}.
\end{equation}
By comparison with Eqn. \ref{agas}, we finally arrive at:
\begin{equation}
\alpha_\mathrm{gas} = ap_\mathrm{gas}\,n_g\,\frac{\iint_\mathrm{gas}n_0\,|\mathbf{E}|^2\,dx\,dy}{\iint_{\mathbb{R}^2}n^2\,|\mathbf{E}|^2\,dx\,dy} = an_g n_0^{-1} \Gamma_\mathrm{E}\,p_\mathrm{gas},
\end{equation}
where $\Gamma_\mathrm{E}$ is the electric field confinement factor:
\begin{equation}
\Gamma_\mathrm{E}\equiv\frac{\iint_\mathrm{gas}n_0^2\,|\mathbf{E}|^2\,dx\,dy}{\iint_{\mathbb{R}^2}n^2\,|\mathbf{E}|^2\,dx\,dy}.
\end{equation}
We emphasize that this is not, in general, equal to the ``traditional'' confinement factor $\Gamma$ defined based on the proportion of modal Poynting vector in gain/loss region, as has been previously noted\cite{gain}.

The remainder of our gas sensitivity analysis follows that of Siebert et al.\cite{sensoranalysis}, in which an ideal gas-light interaction wavelength is found as a function of the gas concentration where maximum sensitivity to deviations is desired. For gas detection applications, maximum sensitivity at zero concentration is desired, and the ideal interaction waveguide length is simply found to be $l_\mathrm{opt}=\alpha_\mathrm{base}^{-1}$. This then leads directly to Main Eqn. 1.

\section{Optical absorption of graphene for longer wavelengths}
In Supplementary Figure \ref{grcond}, we apply Main Eqn. 3 to predict how the real part of our graphene's optical conductivity (and thus optical absorption) would change at longer wavelengths using carrier concentration spread and Drude scattering time material parameters extracted from our device model. We find that the optical absorption remains roughly the same if not even higher at longer wavelengths, which is due to increased intraband absorption; therefore we anticipate similar device performance can be achieved at these wavelengths.

\bibliographystyle{ieeetr}
\bibliography{wimp}% Produces the bibliography via BibTeX.